\documentclass[aps, prl, twocolumn, superscriptaddress, amsmath,  tightenlines, longbibliography]{revtex4-1}

\usepackage{graphicx}
\usepackage{dcolumn}
\usepackage{bm}
\usepackage{braket}
\usepackage{float}	
\usepackage{chapterbib}
\usepackage{amsmath}

\begin{document}
\preprint{APS/123-QED}
\title{Direct observation of energy band attraction effect in non-Hermitian systems}
\author{Maopeng Wu}
	\affiliation{ Department of Engineering Physics, \\
			Tsinghua University, Beijing 100084, China}
	\affiliation{State Key Laboratory of Tribology, Department of Mechanical Engineering,\\
Tsinghua University, Beijing 100084, China}
\author{Ruiguang Peng}
	\affiliation{State Key Laboratory of Tribology, Department of Mechanical Engineering,\\
Tsinghua University, Beijing 100084, China}
\author{Jingquan Liu}
	\email{jingquan@tsinghua.edu.cn}
	\affiliation{ Department of Engineering Physics, \\
			Tsinghua University, Beijing 100084, China}
	
\author{Qian Zhao}%
 	\email{zhaoqian@tsinghua.edu.cn}
 	\affiliation{State Key Laboratory of Tribology, Department of Mechanical Engineering,\\
Tsinghua University, Beijing 100084, China}
\author{Ji Zhou}
	\email{zhouji@tsinghua.edu.cn}
	\affiliation{ State Key Laboratory of New Ceramics and Fine Processing,\\
School of Materials Science and Engineering, Tsinghua University, Beijing 100084, China}
\date{\today}

\begin{abstract}
The energy band attraction (EBA) caused by the non-orthogonal eigenvectors is a unique phenomenon in the non-Hermitian (NH) system.  However, restricted by the required tight-binding approximation and meticulously engineered complex potentials, such effect has not been experimentally demonstrated. Here, an experimentally verifiable model is proposed based on the photonic counterpart of the all-dielectric Mie-resonator lattice in a parallel-plate transmission line. Through theoretical derivation, we directly connect the transmission spectra with eigenvalues and eigenvectors of the NH Hamiltonians.  By precisely tuning the resonance loss of the Mie-resonators, the evolution of the EBA effect in two-level NH systems, from gapped bands, gapless bands to flat bands, is directly observed for the first time. Furthermore, such effect can be extended to a graphene-like two-dimensional NH system. Our works show a metamaterial approach towards NH topological photonics and offer a deeper understanding of band theory in open systems.
\end{abstract}
\maketitle
\textit{Introduction}.---Systems with non-Hermitian(NH) Hamiltonians, usually induced by complex potential or asymmetrical hopping, have received increasing attention in many branches of physics and optics. Non-Hermitian systems not only promise an extra flavor to expand the parameter space of physics into the full complex domain but also exhibit singular phenomena. For example, NH operators that have PT-symmetry exhibit purely real spectra\cite{bender1998real,guo2009observation}; Non-Hermitian energy band theory has a nontrivial topology\cite{shen2018topological,lee2016anomalous,leykam2017edge,kunst2018biorthogonal,Poli2015Selective,Rudner2009Topological,Zeuner2015Observation,weimann2017topologically,weimann2017topologically};  Open boundary has significant differences with the periodic boundary and exhibits NH skin effect\cite{Yao2018Edge,Yao2018Non,song2019non,yokomizo2019non,longhi2020non,luo2019higher}; PT-symmetry periodic potentials induce optical solitons\cite{wimmer2015observation,muniz20192d}; A new class of degeneracies(dubbed as exceptional points)lead to unidirectional transmission or reflection\cite{regensburger2012parity,lin2011unidirectional,peng2014parity,feng2011nonreciprocal}, topological half charge\cite{lee2016anomalous,hassan2017dynamically}, and Fermi arc\cite{zhou2018observation}. Besides, those singular phenomena also offer some practical device applications, such as the laser with single-mode\cite{hodaei2014parity}, the microlaser with orbital angular momentum\cite{miao2016orbital}, and the sensing with enhanced sensitivity\cite{wiersig2014enhancing,chen2017exceptional,hodaei2017enhanced}.\par
Energy band repellence is an essential concept in the energy band theory, namely, adjacent energy levels that interact under the influence of perturbation always repel each other, give rise to an upward shift of higher level, and a downward shift of lower level, respectively. However, this concept is no more valid in NH systems and recent experiments have also observed attractive level crossings in two coupled systems, such as the Fabry-Perot-like cavity\cite{harder2018level}, coplanar-waveguide-based resonator structures\cite{bhoi2019abnormal,yang2019control}, and cavity-magnon polariton\cite{grigoryan2018synchronized,boventer2020control}. Two coupled harmonic oscillators model with NH dissipative coupling terms have been proposed to interpret the experiments\cite{yao2019microscopic,yu2019prediction}. Recent works also show that the energy band attraction(EBA) effect necessitates a redefinition of topological invariants\cite{hasan2010colloquium,qi2011topological,liang2013topological} and results in Wannier-Stark ladder coalescence and chiral Zener tunneling\cite{longhi2020non}. \par 
However, experimental and theoretical exploration of EBA has so far mostly dealt with two coupled-mode systems. This unique NH feature can be generalized to a wide variety of many-body systems. But, the general implications of these results based solely on a simple two coupled-mode systems remain to be expanded. What's more, the experimental demonstration of more fundamental NH band theory has been hindered by the complicated configuration of tight-binding model and complex potential. Fortunately, topological photonics is opening new perspectives in the study of topological effects of broad interest for quantum condensed matter physics, particularly for NH systems since gain and loss are much more common than electrons in solids\cite{lu2014topological,ozawa2019topological}.\par
In this letter, from the energy band theory perspective, we consider the generalization of the EBA effects in the presence of NH perturbation. We propose a photonic system based on coupled the Mie resonator and theoretically demonstrate the correspondence of eigenvectors and eigenvalues between the NH system and this photonic system, where the energy and wavefunctions of NH system are corresponding to the resonant frequency and magnetic distributions of photonic system. With the help of visual microwave measurements and fast-growing metamaterial methods, for the first time, EBA effect and its evolution process in one-dimensional(1D) NH system are experimentally observed, that is, two bands of microwave resonance bipartite chain gradually coalesce into a flat band by precisely tuning the NH parameter. Furthermore, our theory indicates that such EBA effect is able to be extended to a two-dimension(2D) system with graphene-like Mie resonance lattice, and the theoretical calculation results are greatly coincident with the simulations.

\begin{figure}
\includegraphics[width=9cm]{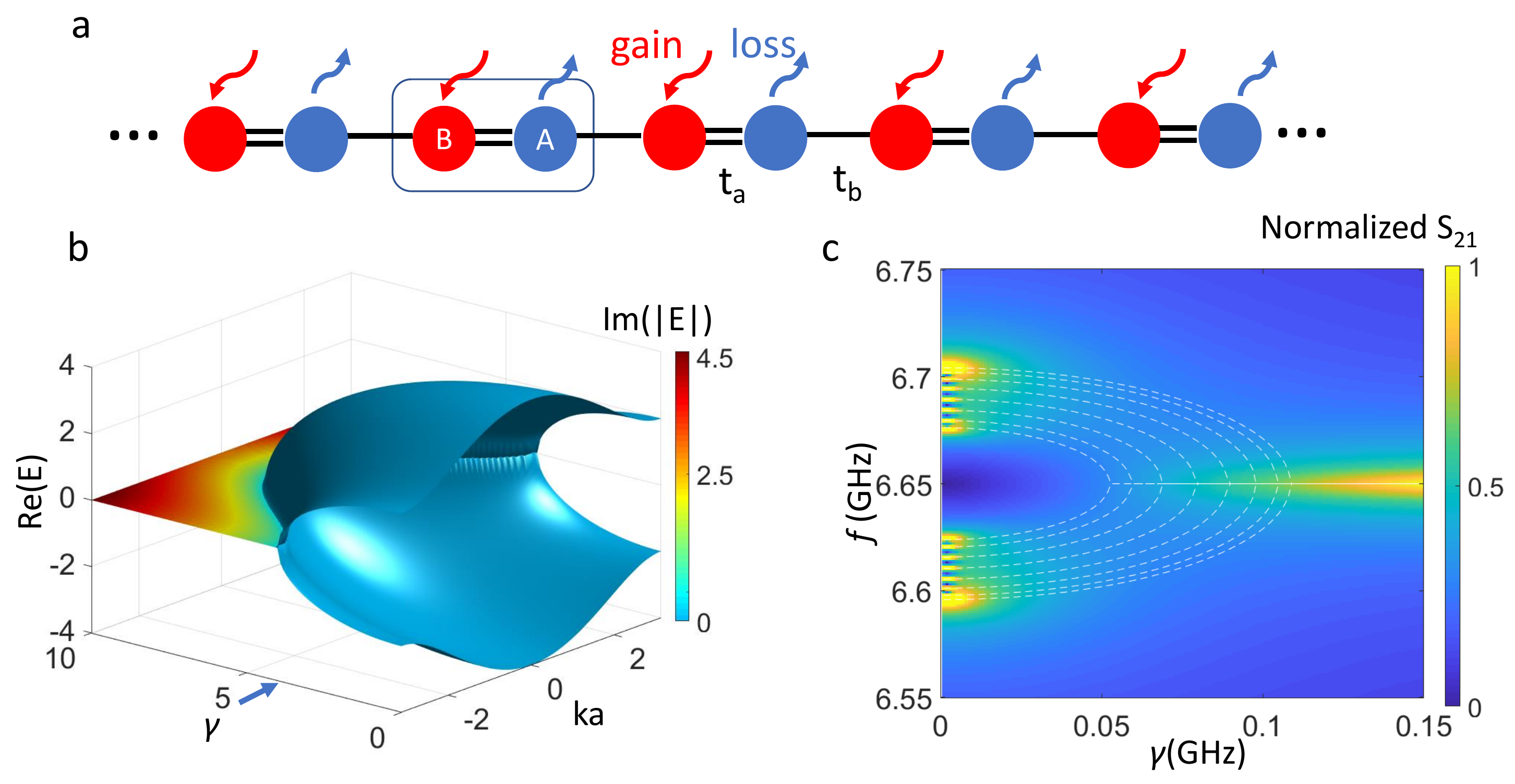}
\caption{\label{fig1} Non-Hermitian SSH model and energy band attraction effect in quantum systems. (a) Lattice representation of the non-Hermitian model in Eq.(1).  All sites in blue (red) have particle loss (gain) with a rate of $\gamma$. The dotted box indicates the unit cell and single bar (parallel bars) indicates the weak (strong) coupling coefficients of the nearest neighbors. (b)Band structure of NH SSH model as functions of $\gamma$, varies in [0,10]; $t_b=\frac{1}{3}t_a=1$. The transition point($\gamma=4 $) between the PT-symmetry phase and PT-symmetry spontaneously broken phase is indicated by an arrow. (c) The transmission spectra $S_{21}$ as a function of NH parameter $\gamma$ calculated from Eq.6. The color scale describes the transmission amplitude, and the white dashed lines indicate the real part of eigenvalues of system Hamiltonian $H$. PT-symmetric region: $\gamma \in  \left [ 0, 0.053\right)$; Partially PT-symmetry broken region: $\gamma \in  \left [ 0.053, 0.109\right)$; Fully PT-symmetry broken region: $\gamma \in \left [ 0.109, \infty \right)$.}
\end{figure}
\textit{Energy band attraction effect}.---We would like to start from a general 1D NH Su-Schrieffer-Heeger(SSH) model (see Fig.\ref{fig1}(a)) to explain EBA theoretically. Similar models are relevant to quite a few experiments to study NH bound states\cite{Poli2015Selective,Zeuner2015Observation,weimann2017topologically}. Different from the traditional SSH model, equivalent positive and negative imaginary contribution to the potential terms are added on the selected sites, where imaginary parts of complex potential can be understood as the particles gain/loss. The following tight binding equation describe the Hamiltonian of the resulting system in the Bloch representation as the function of the lattice constant $a$, Brillouin vector $k$, and Pauli matrix $\boldsymbol{\sigma }$.
\begin{equation}
H\left ( k \right )=\left ( t_a+t_b coska\right )\sigma _x+t_bsinka\sigma_y+i\gamma\sigma_z
\end{equation}
where $t_a$ and $t_b$ denote the intra dimer and inter dimer coupling strengths. Non-Hermitian term $\gamma$ indicates the gain/loss within one dimer. The Hamiltonian here posses PT-symmetry $\sigma_xH\left ( k\right )^*\sigma_x=H\left ( k\right )$ and pseudo-anti-Hermiticity $\sigma_zH\left ( k\right )^\dagger \sigma_z=-H\left ( k\right )$. The former symmetry guarantees all the eigenvalues are real under the PT-symmetry phase\cite{bender1998real}. The latter one gives rise to pairwise eigenvalues and promises nontrivial topology\cite{lieu2018topological,takata2018photonic}. Energy dispersion of $H\left ( k\right )$ can be calculated as follow
\begin{equation}
E\left ( k\right )=\pm \sqrt{t_a^2+t_b^2+2t_at_bcos\left ( ka \right )-\gamma^2}
\end{equation}
The dispersion sensitively depends on the value of the NH term $\gamma$, as shown in Fig.\ref{fig1}(b). Unlike energy band repellence of Hermitian systems, the two-level gapped bands firstly go close, then coalesce, and ultimately  merge into a flat band as $\gamma$ increases, which is independent of Brillouin vector $k$. While in the Hermitian energy band repellence case, correct tuning of the gapped system parameters reduces the bandgap and causes the degeneracy at the threshold, but continuous tuning will make the gapless bands reopen again. For example, inversion symmetry-breaking Semenoff mass term of honeycomb lattice Hamiltonian plays an essential role in the Valley Hall effect, where correct tuning this mass term will close the bandgap at Dirac points. While continuing to tuning will reopen the bandgap that belongs to different topological phases\cite{wu}. Notably, the EBA effect still exists in the loss-dominant systems, which are the standard experimental construction to achieve passive PT-symmetry\cite{Poli2015Selective,Zeuner2015Observation,weimann2017topologically}. \par

\begin{figure}[H]
\includegraphics[width=8cm]{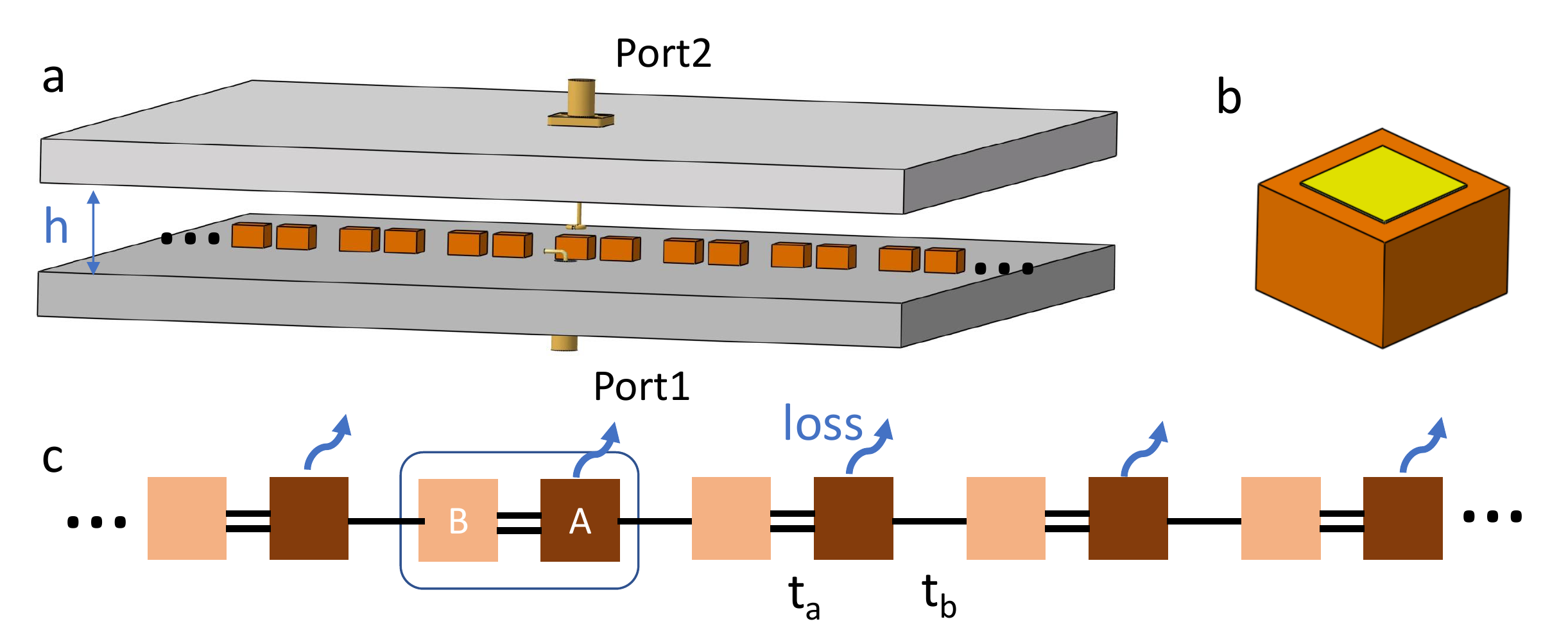}
\caption{\label{fig2}The Schematic of experimental analogue and the passive PT-symmetry Mie resonators lattices. (a)The Mie resonance lattice  in the parallel-plate transmission line. The cutoff frequency of the parallel-plate transmission line, determined by separate distance $h$, is lower than the Mie resonance frequency to ensure resonance coupling. (b)Absorption implementation. Absorption is introduced on selected sites by gold sputter deposition on top of the resonators and can be controlled by varying the area of sputter deposition(see supplementary). (c)Schematic of the complex SSH chain with staggered resonance coupling amplitudes $t_a$ and $t_b$. The coupling amplitudes can be controlled by varying the resonator spacings(see supplementary). 
}
\end{figure}
\textit{Proposals for the microwave analogue.}---Next, we will elaborate on how we bridge the gap between NH band theory and the experimental realization. For theoretical model, as illustrated in the Fig.\ref{fig2}(a), two insulated conductors constitute the parallel-plate transmission line, in which a set of coupled dielectric resonators are placed. The first-order Mie magnetic-dipole-resonance of resonators is excited purposely below the cutoff frequency of TM modes in a parallel-plate transmission line. Thus electromagnetic field is mostly confined within the resonators and spreads out evanescently. The weak coupling induced by the evanescent field ensures electromagnetic traveling between two adjacent resonators and can be controlled by adjusting the separation distance\cite{wu}. Kink antenna and shielded loop antenna, namely the port one and the port two, are set aside the resonators lattice and coupling between them can be neglected.\par
If we collect the excitation amplitudes $a_{1,2,3...n}$of resonators $1,2,3...n$ in a vector $\ket{a\left (t \right )}=\left [ a_1\left ( t\right ) ,a_2\left ( t\right ),...a_n\left ( t\right )\right ]^T$, where n enumerates the discs, and input and output amplitudes at the ports in $\ket{s_+\left ( t\right )}=\left [ s_+^1\left ( t\right ),s_+^2\left ( t\right )\right ]$ and $\ket{s_-\left ( t\right )}=\left [ s_-^1\left ( t\right ),s_-^2\left ( t\right )\right ]$ respectively, the resulting system evolves in time $t$ described by\cite{suh2004temporal} 
\begin{align}
\frac{d}{dt}\ket{a}&=\left ( jH-\Gamma \right )\ket{a}+K^T\ket{s_+\left ( t\right )} \\ 
\ket{s_{-}\left ( t \right )}&=\ket{s_+\left ( t\right )}+K\ket{a}
\end{align}
where $H$ is the Hamiltonian of the resonators lattice. $K$ is the coupling matrix between ports and resonators lattice whose elements are given by $K_{n,m}=M\left ( w\right )\varphi _n^*\left ( r_m\right )$\cite{wu}, where $M(w)$ is the parameter related to antenna structure, $\varphi _n\left ( r_m\right )$ is the eigenfunction of $H$, and $r_m$ denotes the port position. $\Gamma^{ant}$ is the decay rate of the ideal lossless resonators lattice contributed to fields decay into ports and satisfies $2\Gamma^{ant}=K^TK$\cite{suh2004temporal}. For externally incident excitation $\ket{s_+}$ at frequency $w$, we can write the scattering matrix $S$ in the frequency domain as
\begin{equation}
S=\frac{\ket{s_-}}{\ket{s_+}}=I+\frac{KK^T}{i\left ( w -H\right )+\Gamma^{ant}}
\label{s_parameter}
\end{equation}
The frequency range used in experiments is of the order of or less than $300$MHz\cite{bellec2013tight,bellec2013topological}, thus $M_n\left ( w\right )$ can be assumed to be nearly constant. Above all transmission through the two ports at position $r$ and ${r}'$ reads
\begin{equation}
\label{s_21}
S_{2,1}=M_1M_2\sum_{n}\frac{\phi_n\left ( r\right )\varphi _n^*\left ( {r}' \right )}{i\left ( w-w_n^{real}\right )+\Gamma_n^{imag}+\Gamma^{ant}}
\end{equation}
where, $w_n^{real}$($\Gamma_n^{imag}$) is the real(imaginary) part of eigenvalues of $H$.  $\phi _n\left ( r_m\right )$ is the eigenfunction of $H^{\dagger}$\par
Equation.\ref{s_21} describes the transmission spectra in terms of the eigenvectors and eigenvalues of the system's Hamiltonian. It suggests that each resonance frequency of transmission spectra is characterized by the real part of eigenvalues $w_n^{real}$, whereas the imaginary part represents the linewidth of the resonance. If we fix the first port position, the magnitudes of the resonances corresponding to given eigenfrequencies (e.g.,$w_1^{real}$) will be mostly proportional to the eigenvectors $\phi_n\left ( r\right )$ at port two position $r$, that is $ S_{2,1} \propto \phi_n\left ( r\right ) $. The visualization of the eigenvectors distribution associated with each eigenfrequency thus becomes accessible. \par
The standard photonic crystal theory\cite{meade1995photonic} can not explain microwave analogue we proposed here. The Nearly-Free-Electron Approximation and the Tight-binding Approximation are the most commonly simple models employed in the context of condensed matter. They treat the problems in two opposite limits:  either the electrons are considered nearly free and atomic potentials weak, or they are assumed to be bound to atoms, and hopping is treated as a perturbation. The optical analog of the former approximation is the photonic crystal. However, our experimental microwave analog is based on the second one. This difference in physical models greatly facilitates the introduction and controllability of non-Hermiticity through material gain or loss rather than global radiation loss\cite{zhen2015spawning}. \par
\begin{figure}
\includegraphics[width=9cm]{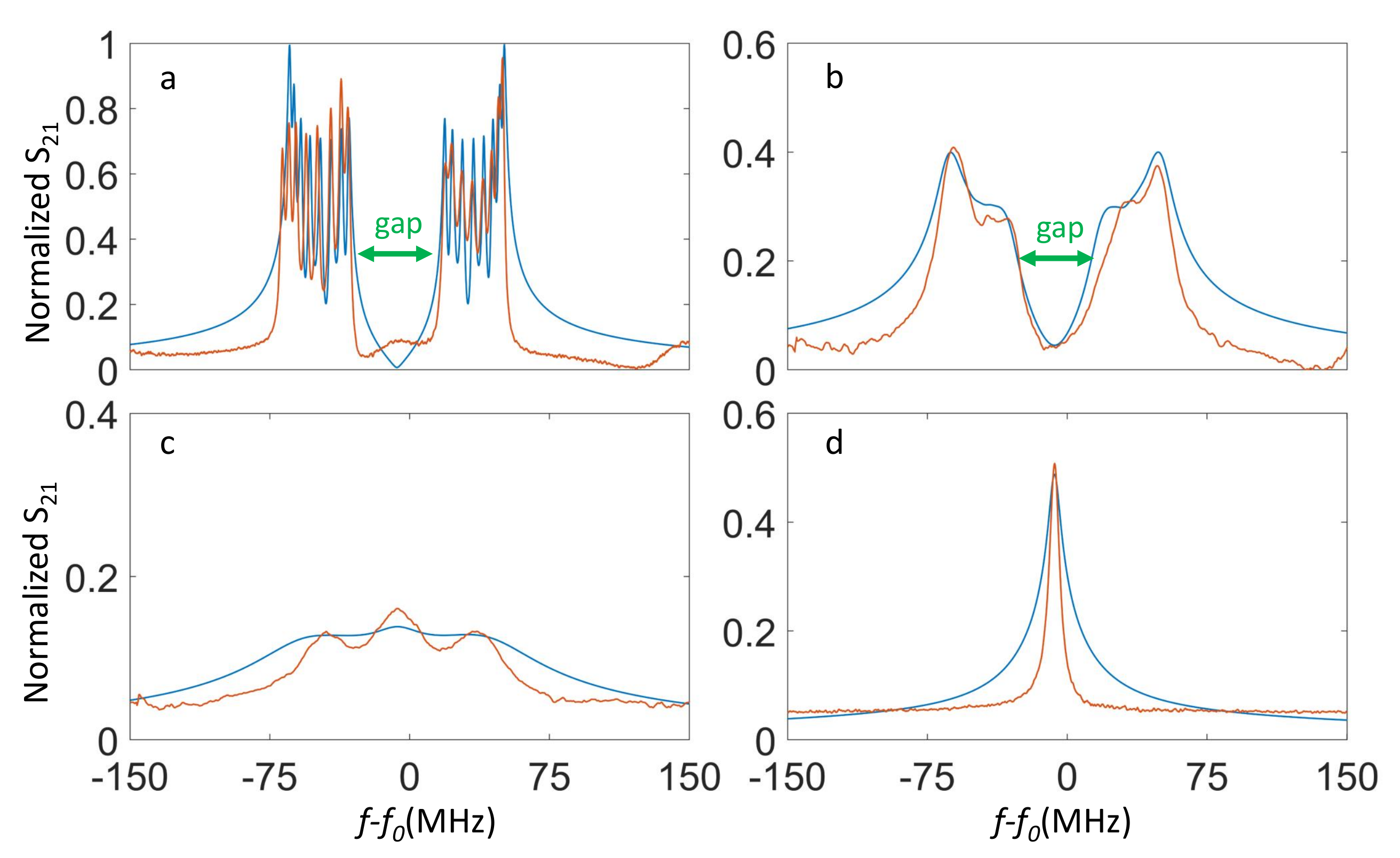}
\caption{Observation of energy bands attraction as NH term $\gamma$ varying through $S$ parameter measurements. The blue lines (orange lines) indicate theoretical prediction (experimentally measured transmission). The lattice is composed of 16 identical coupled dielectric cuboid resonators, which are $5.000$mm height, $7.350$mm width, $7.350$mm length, and a refractive index of $6.50$(the dielectric loss can be neglected). Accordingly, the first-order Mie magnetic-dipole-resonance frequency of the isolated resonator is $f_0=6.493$GHz. All resonators were pre-characterized to obtain a tolerance of below $4.0$MHz. The resonators are placed at alternating distances $d_1=11.50$mm for $t_a=41.0$MHz and $d_2=14.50$mm for $t_b=15.0$MHz. Two conduct plates are separated by $h=17.00$mm corresponds to the cutoff frequency $8.75$GHz. (a)$\gamma=0$, Hermitian case. In this case, all the eigenvalues are real, each resonance peak corresponds to the eigenvalue of the Mie resonance lattice. (b)$\gamma\approx 15$MHz belongs to PT-symmetry region. Although all the eigenvalues are real, the resonance peaks are indistinguishable since our experiments are loss-dominant. (c)$\gamma\approx  60$MHz belongs to partially PT-symmetry broken region. The gapped bands coalesce and gap is closed. (d)$\gamma\approx 121$MHz belongs to fully PT-symmetry broken region. All the eigenvalues are complex with identical real parts, which corresponds to a single peak.}
\label{fig3}
\end{figure}
\emph{Results and influence on topology invariance.}---To demonstrate the EBA effect above, we arrange a bipartite chain of resonators according to the NH SSH lattice where absorption is introduced on selected sites (Fig\ref{fig2}.b and c), and calculate the normalized transmission spectra (Fig\ref{fig1}.c). Each magnitude of the transmission spectra corresponding to the eigenvalue is associated with the eigenvectors of $H$, so we average them over all the site positions to obtain the normalized one. Three distinct regions can be defined according to the eigenvalues: PT-symmetric region, all the eigenvalues are real; Partially PT-symmetry broken region, some eigenvalues are real, and the others are complex; Fully PT-symmetry broken region, all the eigenvalues are complex and share identical real parts. Note that the resonators are in cuboid shape purposefully to gain the superiorities in both the simplicity and the flexibility of tunable response\cite{zhao2008experimental,peng2017temperature}. Details of transmission spectra at some specific $\gamma$ are depicted in Fig.3, where experimentally measured results are also plotted. One can see the attraction process from separated resonances peak to a single peak as increasing $\gamma$.\par
 \begin{figure}
\includegraphics[width=9cm]{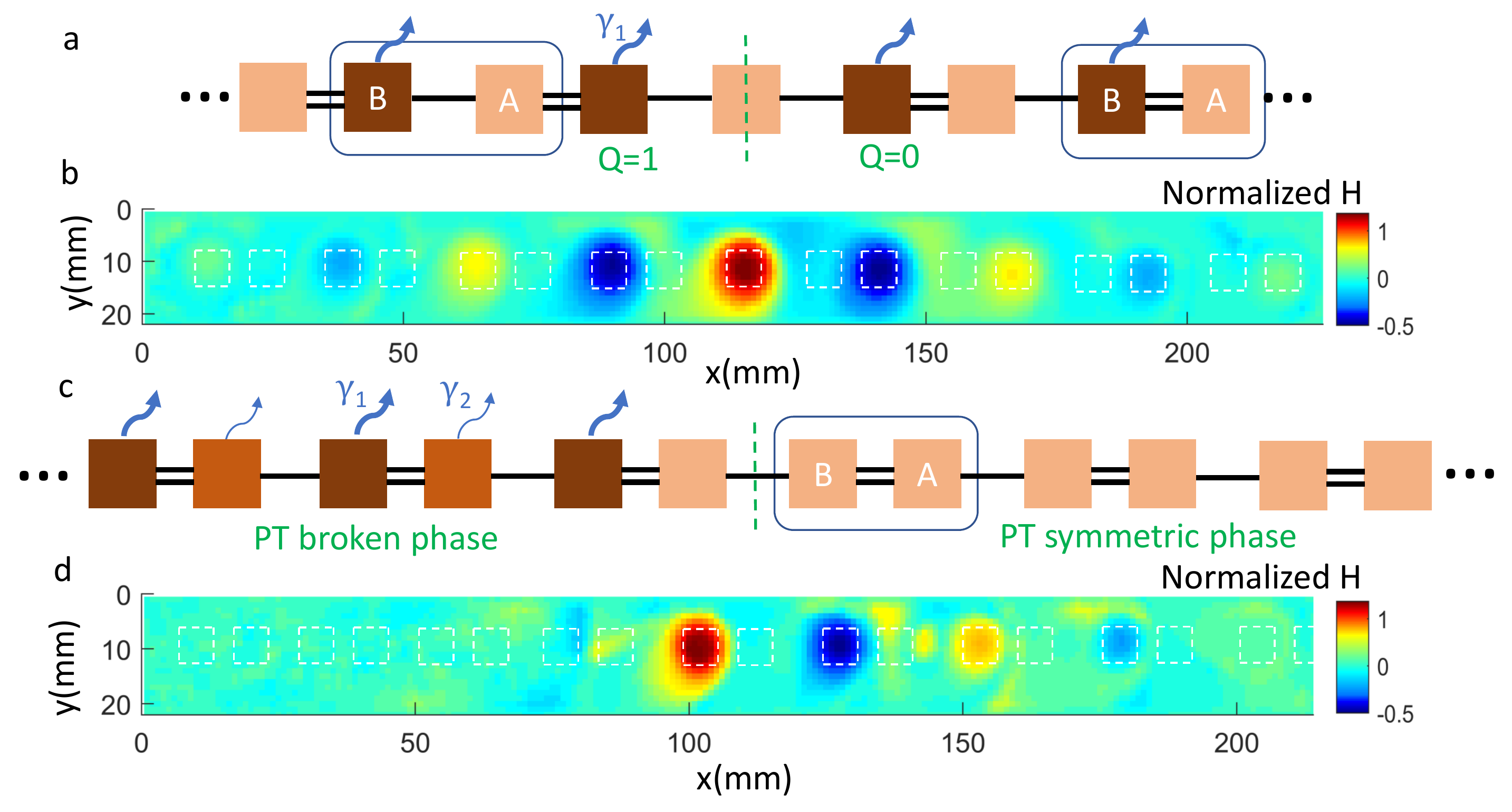}
\caption{\label{fig4}Experimentally detected magnetic field distribution of edge states at bare frequency $f_0=6.493$GHz. (a) and (c) Resonators configuration that supports zero-energy edge modes. (b) and (d) Normalized experimental field distribution corresponds to (a) and (c), and the white dashed lines indicate the contours of resonators.  Loss strength $\gamma_1=121$MHz, $\gamma_1=15$MHz, and coupling strength $t_1=41.0$MHz,  $t_2=15.0$MHz. Q stands for Global Berry phase. (b)Magnetic field of the edge state is exponentially localized around the topological interface and has nonvanishing components only on the A sublattice. (d)The edge state also exits around the PT interface even the systems is topological trivial.}
\end{figure}
Furthermore, we observed the topological edge states using our experimental systems (Fig.\ref{fig4}). In the band theory framework, bulk topological invariance plays an essential role in topological materials.  According to the principle of bulk-boundary correspondence,  the invariance can accurately predict the existence of gapless edge modes,  e.g., in 2D quantum Hall theory, a given band has a quantized Chern number, and the difference in Chern number across the interface dictates the number of the gapless edge modes\cite{hasan2010colloquium,qi2011topological}. If we simply extend traditional topological materials theory into the NH framework, the topological invariance can be a complex number\cite{liang2013topological}, which implies the sudden breakdown and loss of its physics meaning, e.g., in our NH SSH model, the complex onsite potential destroys the quantization and the reality of the Berry phase. But, interestingly, the summation of the complex Berry phase, namely Global berry phase, is quantized and has been demonstrated to serve as the topological invariance\cite{liang2013topological,lieu2018topological,takata2018photonic}.\par Since EBA occurs and band indices lose their meaning, it's not surprising that topological invariants in the NH model are not a property of the band, but rather of the entire Hamiltonian\cite{lieu2018topological}. Figure.\ref{fig4}(a) shows the realization of the NH resonators lattice interface between two structures that have different Global Berry phase to support topological edge states. Figure.\ref{fig4}(b) shows the experimental field visualization at the bare frequency and reveals that the field is localized exponentially on the interface. Figure.\ref{fig4}(c) and (d) also illustrates another realization of the NH interface that can support the zero-energy edge state, despite the trivial Global Berry phase in nature. This new type of edge state is driven by the PT phase transition and characterize robustness\cite{pan2018photonic}. It is worth noting that not only the amplitude of the field but also the phase are detected in our setup, which can not reach easier by most of the other experiment systems.\par   
 \begin{figure}
\includegraphics[width=9cm]{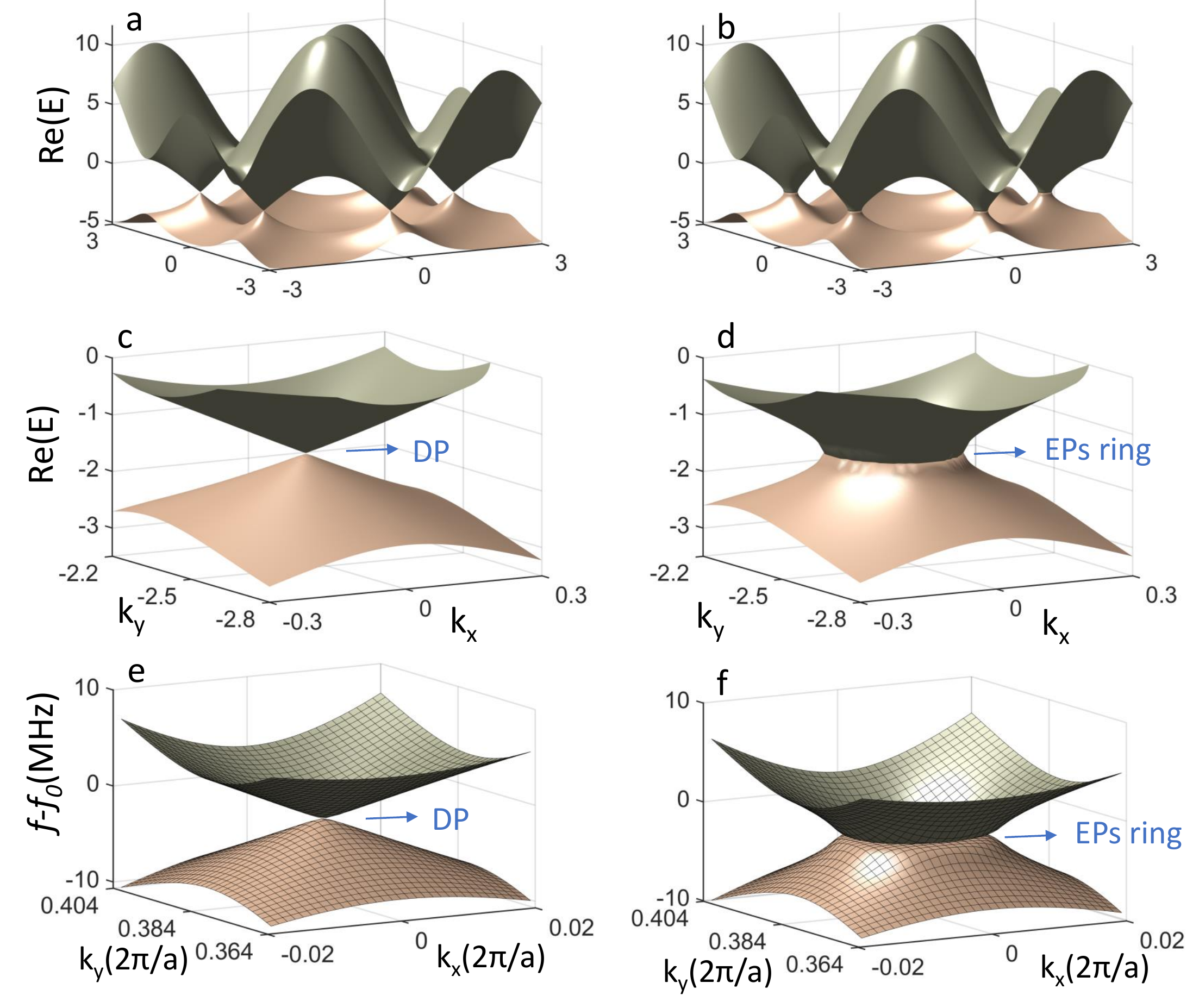}
\caption{\label{fig5}Theoretical and numerically energy band structure in Hermitian (Left panel) and non-Hermitian (right panel) honeycomb lattice.  Only the real parts of the eigenvalues are plotted. (a)-(d), Theoretical results of NH graphene, with $t_1=2.7$eV, $t_2=-0.54$eV, $t_3=0.1$eV and $\gamma=0.5$eV. The analysis predicts that the real parts of the eigenvalues stay as a constant inside the EPs ring, indicating flat bands in dispersion. (e) and (f), simulation results of Hermitian and NH resonances lattice, with $t_1=16$MHz, $t_2=-1.45$MHz, $t_3=1$MHz, $\gamma=21$MHz and lattice constant $a=15$mm(see supplementary). Arrows indicate Dirac point (DP) and the ring of exceptional points(EPs ring).  }
\end{figure}
\textit{Graphene-like structure.}---At last, we further generalize the EBA effect into 2D systems and numerically demonstrate it using the theoretical framework we proposed.  As an illustration, we study a concrete lattice model--NH 2D graphene, which has honeycomb lattice. One of the most exciting aspects of the graphene is the existence of linear crossing, the so-called Dirac cone between the first and the second band, where intersection points are dubbed as Dirac points\cite{neto2009electronic}. With the nearest-neighbor, next-nearest-neighbor, and third-nearest-neighbor into consideration, the chiral symmetry is broken. Therefore, the band is asymmetric (the full band structure Fig.5(a) and a zoom-in around the Dirac region Fig5(c)). Figure.\ref{fig5} (b) and (d) show the band structure of NH graphene, where alternating gain and loss are added to the sublattice\cite{wu}. When the NH parameter is present, the EBA effect will deform Dirac cone, turning Dirac point into a ring of exceptional points and turning linear dispersion into exponential one. This novel dispersion can potentially serve as hyperbolic metamaterials\cite{poddubny2013hyperbolic}. Accordingly, Hermitian and NH simulation results of Mie resonators arranged in honeycomb lattice are shown in Fig.\ref{fig5}(e) and (f), and the dispersion relation is consistent with the analysis. More details about NH graphene analysis and simulation configuration are given in supplementary\cite{wu}.

\textit{Conclusion}---In conclusion, we have predicted the energy band attraction effect in the non-Hermitian system and explain it through the analytic solutions of the NH SSH model and the NH graphene model. A new metamaterial approach based on all-dielectric coupled Mie resonators is further proposed to analogize the NH system, and the correspondence is explicitly demonstrated. Owing to the advantage of the accessibility of the band structure and field distribution visualization, we have made a direct observation of the energy band attraction and topological edge states affected by this effect. Our works help to understand the framework of the band theory in NH systems and might serve as a viable platform for further promising applications such as robust light steering\cite{zhao2019non}, flat bands transport without diffraction\cite{biesenthal2019experimental}, and NH hyperbolic metamaterials\cite{poddubny2013hyperbolic,hou2020topological}. The physics and realization presented here can be generalized to a wide variety of non-Hermitian systems, such as 3D NH Weyl semimetals, which will be left for the future.
\begin{acknowledgments}
This work is supported by the Beijing Municipal Science \& Technology Commission (Z191100004819001), the National Natural Science Foundation of China (51872154), the Science and Technology Plan of Shenzhen City (JCYJ20170817162252290), and the Chinese State Key Laboratory of Tribology. The authors gratefully acknowledge discussions with Dr. J. D. Wen, Dr. L. Kang, Dr. N.-H. Shen, Dr. P. Zhang, Prof. Q. Ren.
\end{acknowledgments}
\nocite{*}
\bibliography{reference}
\clearpage

\appendix
\chapter{Supplemental Material}
\section{Energy Level attraction in honeycomb lattices}
\begin{figure}[H]
\center
\includegraphics[height=4cm]{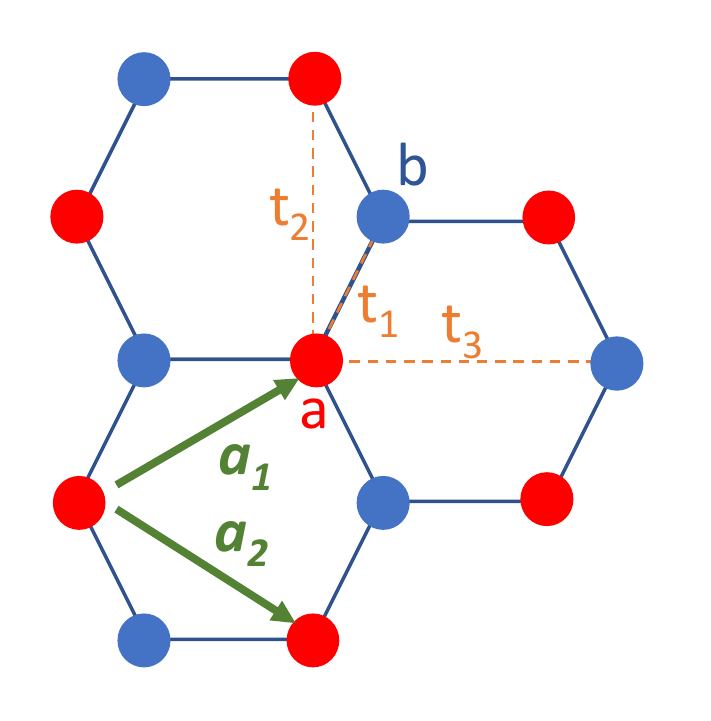}
\caption{\label{fig_appendix1}  Sketch of NH honeycomb lattices. Sublattice a and b exhibit complex potentials，$\alpha +\lambda i$ and $-\alpha -\lambda i$, respectively. Primitive vectors $a_1$ and $a_2$ are indicated by the green arrows. }
\end{figure}

\begin{figure*}
\includegraphics[height=6cm]{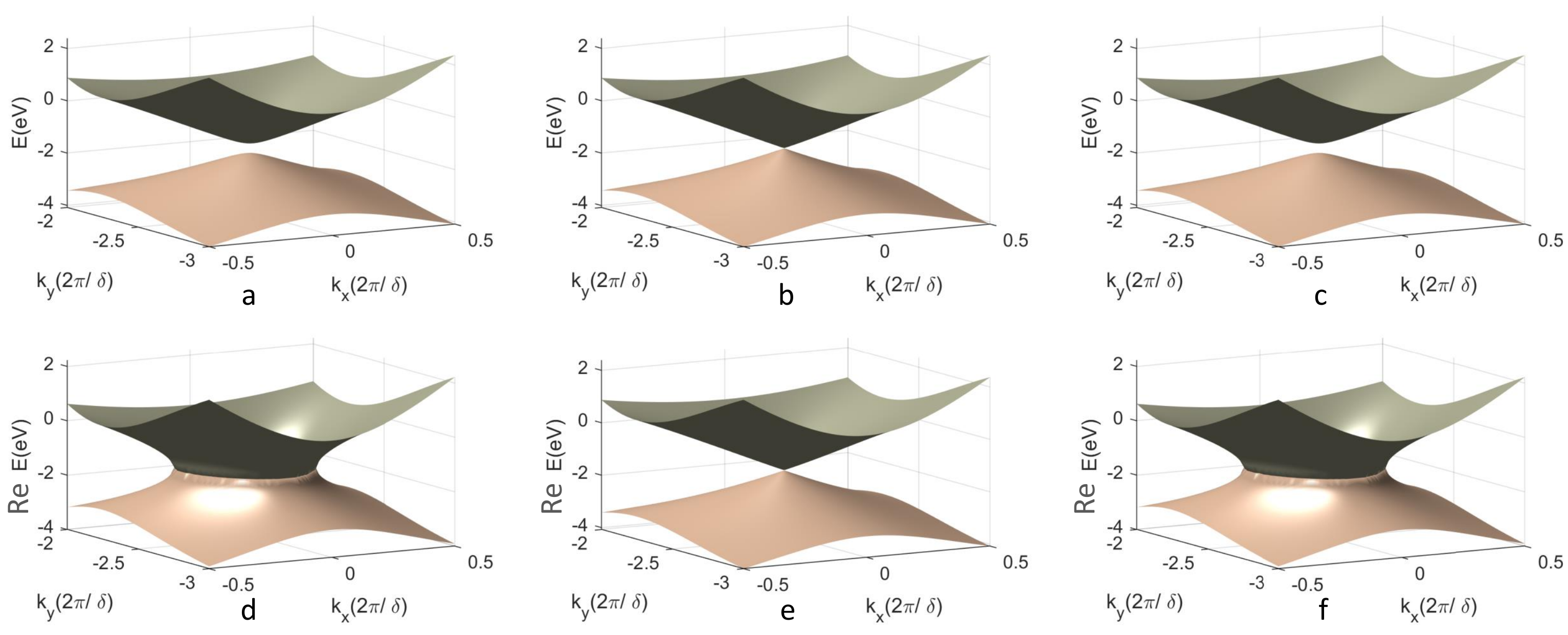}
\caption{\label{fig_appendix2}Band structure around Dirac region with Hermitian and Non-Hermitian perturbation. $t_1=2.7eV$, $t_2=-0.54eV$, $t_3=0.1eV$. Up panel: Hermitian perturbation, and $\alpha=0$. (a)$\lambda=-0.1$. (b)$\lambda=0$. (c)$\lambda=0.1$. Down panel: Non-Hermitian perturbation, and $\lambda=0$. (d)$\alpha=-0.5$. (e)$\alpha=0$. (f)$\alpha=0.5$.  }
\end{figure*}

In this section, we will talk about EBA in 2D NH systems in details and demonstrate that non-Hermitian perturbation can cause the levels to merge. The structure of honeycomb lattices with alternating complex potentials (complex Semenoff mass) is sketched in Fig \ref{fig_appendix1}.  Here, hopping amplitudes between nearest-, next-nearest-, and third-next-nearest neighbor atom are in consideration, which is respectively denoted as $t_1$, $t_2$, and $t_3$(only $t_2$ hops between the same sublattice). The tight-binding Hamiltonian in real space writes
\begin{widetext}
\begin{equation}
H=t_1\sum_{<i,j>}\left ( a_{i}^\dagger b_{j} +h.c\right )+t_2\sum_{<<i,j>>}\left ( a_{i}^\dagger a_{j} +b_{i}^\dagger b_{j}+h.c              \right)+t_3\sum_{<<<i,j>>>}\left ( a_{i}^\dagger b_{j} +h.c\right )+\left ( \alpha +\lambda i\right ) \sum_{i}a_{i}^\dagger a_{i}- \left ( \alpha +\lambda i\right ) \sum_{i}b_{i}^\dagger b_{i}
\end{equation}
\end{widetext}
where $a_{i}^\dagger$ ($a_{i}$) annihilates (creates) a particle on site $i$. $\lambda$ is imaginary on-site sublattice potential which makes the system non-Hermitian. $\alpha$ is the real parts of the complex potentials appeared as an inversion symmetry-breaking Semenoff mass term. The Hamiltonian above can be rewritten in momentum space
\begin{subequations}
\begin{equation}
H\left ( \boldsymbol{k}\right )=\begin{pmatrix}
f_2\left ( \boldsymbol{k}\right )+\alpha +\lambda i &f_1\left ( \boldsymbol{k}\right )+f_3\left ( \boldsymbol{k}\right ) \\ 
f_1^*\left ( \boldsymbol{k}\right )+f_3^*\left ( \boldsymbol{k}\right )  & f_2\left ( \boldsymbol{k}\right ) -\alpha -\lambda i
\end{pmatrix}
\end{equation}
\begin{equation}
f_1\left ( \boldsymbol{k}\right )=-t_1\left ( 1+e^{i\boldsymbol{k} \cdot \boldsymbol{a_1}}+e^{i\boldsymbol{k} \cdot \boldsymbol{a_2}}\right )
\end{equation}
\begin{equation}
f_2\left ( \boldsymbol{k}\right )=-2t_2\left [ cos\left ( \boldsymbol{k}\cdot \boldsymbol{a_1}\right )+ cos\left ( \boldsymbol{k}\cdot \boldsymbol{a_2}\right )   +cos\boldsymbol{k}\cdot \left (\boldsymbol{a_1}-\boldsymbol{a_2} \right )            \right ]
\end{equation}
\begin{equation}
f_3\left ( \boldsymbol{k}\right )=-t_3\left [ e^{i\boldsymbol{k} \cdot\left ( \boldsymbol{a_1}+\boldsymbol{a_2}\right )}+e^{i\boldsymbol{k} \cdot\left ( \boldsymbol{a_1}-\boldsymbol{a_2}\right )}+e^{i\boldsymbol{k} \cdot\left ( \boldsymbol{a_2}-\boldsymbol{a_1}\right )}\right ]
\end{equation}
\end{subequations}
$\boldsymbol{k}=\left ( k_x,k_y\right )$ corresponds to the Bloch wave vector, $\boldsymbol{a_1}$ and $\boldsymbol{a_1}$ are the lattice unit
vectors. The energy bands derived from this Hamiltonian have the form\begin{equation}
E\left ( \boldsymbol{k}\right )=f_2\left ( \boldsymbol{k}\right )\pm \sqrt{\left [ f_1\left ( \boldsymbol{k}\right )+f_3\left ( \boldsymbol{k}\right )\right ]^2+\left ( \alpha +\lambda i\right )^2}
\label{sup_energy_band}
\end{equation}
\textit{Hermitian case: energy band repellence.}---When condition $\gamma=0$ is meet, the Hamiltonian mentioned above degenerates into a Hermitian one. The most striking feature of the band structure of this system is the existence of the so-called Dirac region in the vicinity of the intersection points (the vertices) between the first and the second bands. For simplicity, let's focus on band structure around the Dirac region.  When $\lambda$ is negative, the band structure is gapped. The bandgap will reduce as $\lambda$ increases and will close until $\gamma=0$, where the degeneracy point is usually dubbed as the Dirac point. However, if $\lambda$ continues to increase, the bandgap will reopen and enlarge linearly. The Fig.\ref{fig_appendix2} depicted the dispersion relation around the Dirac region as Hermitian parameter $\lambda$ is perturbed. The above process evolving from the gap closing to the gap reopening always accompanies the topological phase transition, which corresponds to the valley topology insulator\cite{236809}. So, the energy band repellence can ensure that adjacent bands will not collapse due to Hermite perturbation.\par
\textit{Non-Hermitian case: energy band attraction.}---The existence of the imaginary parts of the complex Semenoff mass makes the system lose its Hermiticity. Non-Hermitian Hamiltonian breaks the completeness and orthogonality of the eigenbasis, which guarantees the energy band repellence. Fig.7 also shows dispersion relation around the Dirac region under NH perturbation(for simplicity, we take $\alpha=0$). Forced by the EBA, the Dirac point coalesces to the EPs ring regardless of the sign of $\alpha$.
\section{Coupling between Mie resonators and absorption implementation}
\begin{figure}[H]
\includegraphics[width = 9cm]{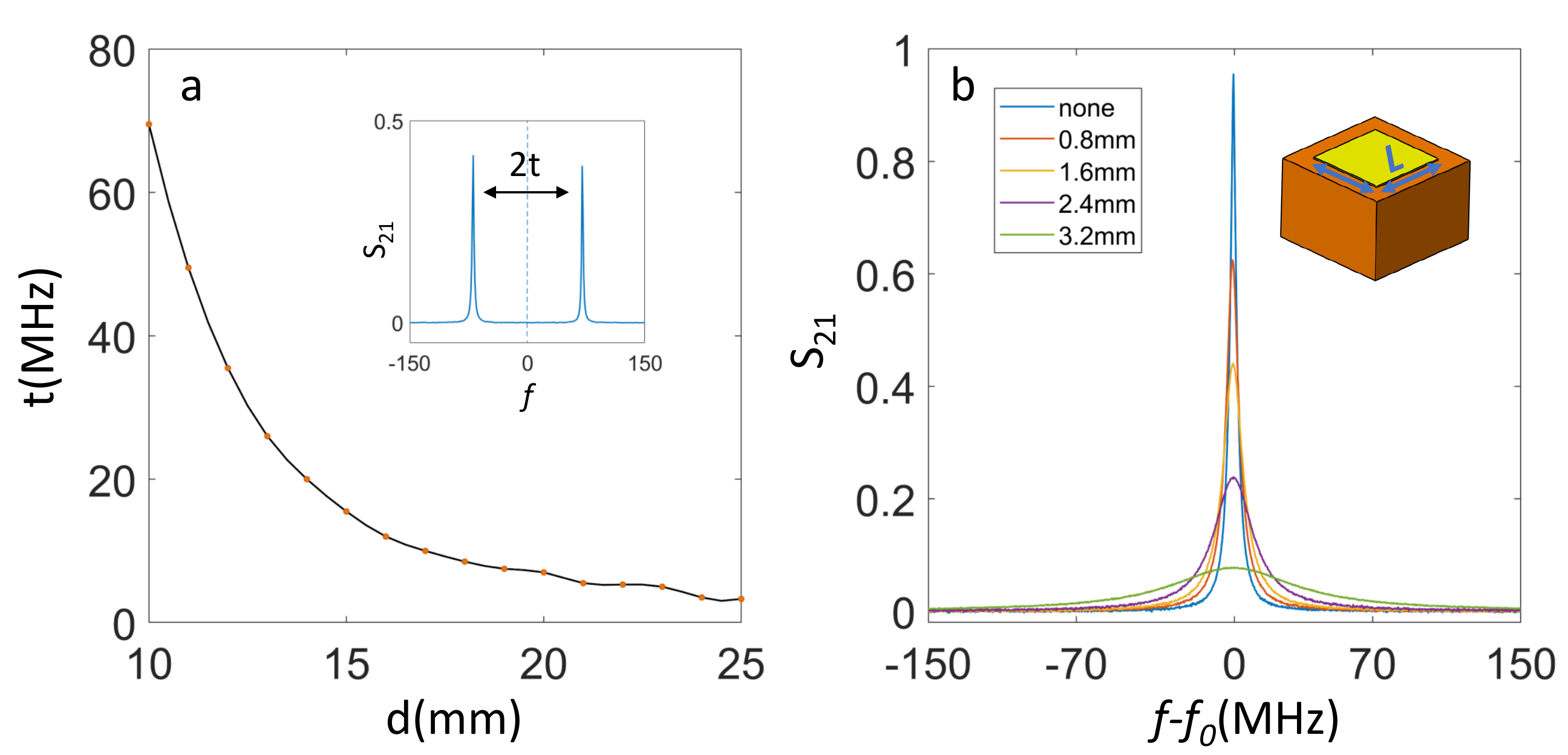}
\caption{\label{sup_fig3}(a)Coupling strength between Mie resonators as separation distance varies and the asymmetric frequency splitting(inset).  (b)Transmission spectra as the area of the gold film varies and absorption implementation(inset). The spectral transmission line has a Lorentzian shape, and linewidth has broadening effects caused by the increasing area of the film. The thickness of the film is about 10nm and the sides length of the film used in the main article are 3mm for $\gamma\approx 15$MHz, 4mm for $\gamma\approx 60$MHz, and 5.6mm for $\gamma\approx 121$MHz.}
\end{figure}

\textit{Coupling}---As we mention in the main article, when Mie resonators are close to each other,  the evanescent field leads to a coupling between them illustrated by asymmetric frequency splitting. This splitting depends on the separation distance $d$ and is twice the coupling strength $t$(see Fig.\ref{sup_fig3}a).\par
\textit{Absorption}---The effect resonator loss is introduced by growing nanometer gold film on the top of the resonators through ion sputtering technology. The resonator loss can be tuned at will by controlling the area or thickness of the film. Figure.\ref{sup_fig3}b shows the transmission spectra as the area of the film varies.

\section{Methods of the simulation}
Numerical calculations are performed using the eigenmode solver in  COMSOL Multiphysics 5.2. We calculate the band structures in a single unit cell configuration accompanied by periodic boundary conditions along the x-direction and the y-direction and PEC boundary conditions along the z-direction. Dielectric loss is added to sublattice b to induce non-Hermiticity.\par
\begin{figure}[H]
\includegraphics[width = 9cm]{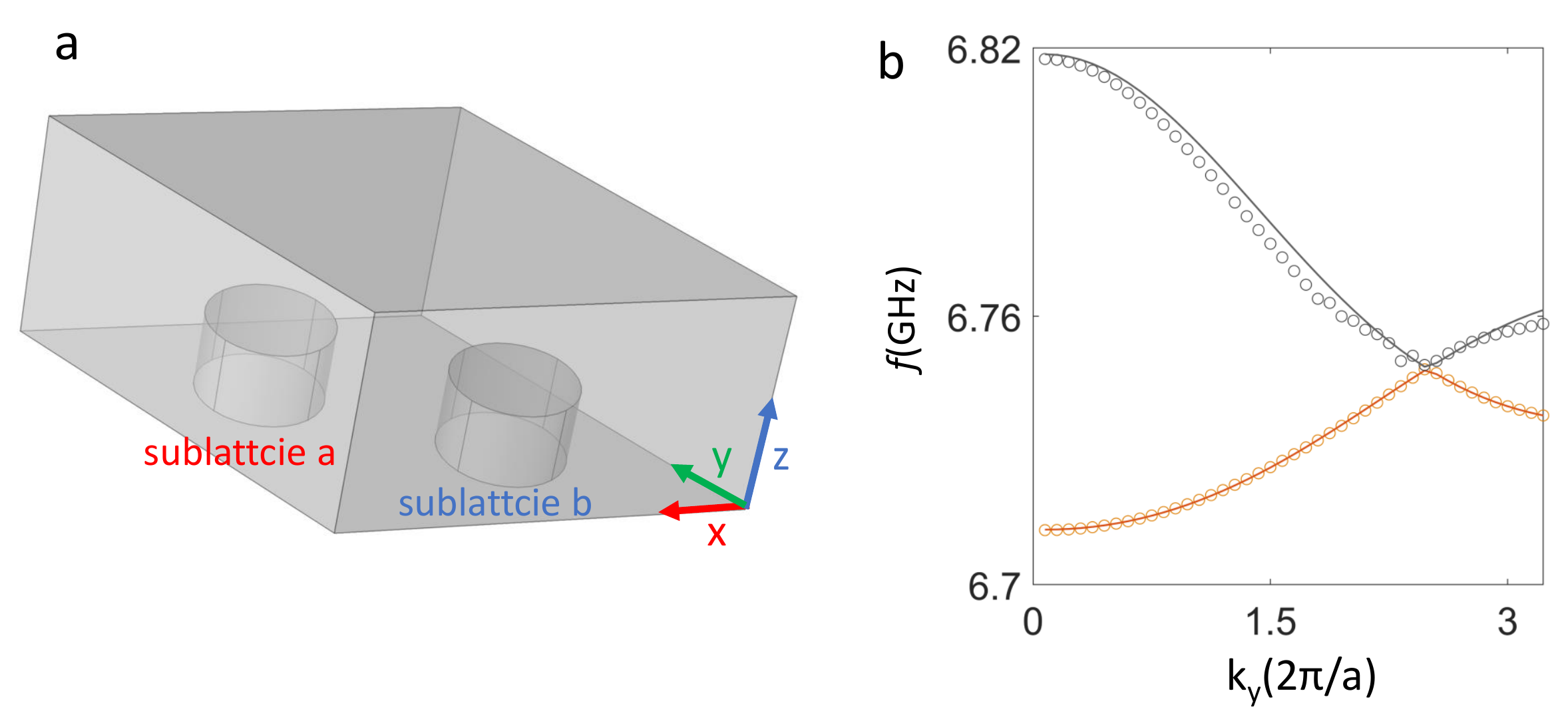}
\caption{\label{sup_fig4}(a)Unit cell of the Mie resonance honeycomb lattice (b)Band structure of the Mie resonance honeycomb lattice along $k_y$. Open circles are numerics of COMSOL calculations. Solid lines are the fit from Eq.\ref{sup_energy_band} with $t_1=16$MHz, $t_2=-1.45$MHz and $t_3=1$MHz.}
\end{figure}

To obtain the coupling amplitudes $t_1$, $t_2$ and $t_3$, we simulate the band structure without  Dielectric loss along the $k_y$ in the Bloch zone and fit the simulation results using Eq.\ref{sup_energy_band}(see Fig.\ref{sup_fig4}). 
\section{$K^T$ in  Temporal Coupled-Mode}
\textit{Total Hamiltonian}---In the presence of interaction V between the eigenvectors of the Mie resonance lattice and states in ports, the total Hamiltonian of the systems in the main article has the form,
\begin{equation}
H_{total}=\begin{bmatrix}
H_{in} & V\\ 
 V^{\dagger }& H_{port}
\end{bmatrix}
\end{equation}
where, $H_{in}$ is the Hamiltonian of the Mie resonance lattice, $H_{port}$ is the Hamiltonian of the ports. let $\Psi=\left [ \psi_{in}\left ( \boldsymbol{r}\right ),\psi^i_{port}\left ( \boldsymbol{r}^i_{port}\right )\right ]^T$ denotes the state of the complete system, where $\psi_{in}\left ( \boldsymbol{r}\right )$ is the lattice wavefunction and $\psi^i_{port}\left ( \boldsymbol{r}^i_{port}\right )$ is the wavefunction of i-th port. According to references\cite{0162052005,19181997}, we can get
\begin{equation}
H_{in}\psi_{in}\left ( \boldsymbol{r}\right )+\sum _{i}T^i_{port}\left ( w\right )\delta \left ( \boldsymbol{r}-\boldsymbol{r}^i_{port}\right )\psi^i_{port}\left ( 0\right )=E\psi_{in}\left ( \boldsymbol{r}\right )
\label{according}
\end{equation}
where, $E$ is the eigenvalue of the total Hamiltonian $H_{total}$. $T^i_{port}$ is the coupling between states of the i-th port and states of the Mie resonance lattice, which is determined by the integral of the current distribution of the antenna. Together with non-Hermitian Green's function in biorthonormal space\cite{2052002,28142002,39442002}, the lattice wavefunction $\psi_{in}\left ( \boldsymbol{r}\right )$ can be written as 
\begin{align}
\label{psilattice}
\psi_{in}\left ( \boldsymbol{r}\right )=\sum _{i}T^i_{port}\left ( w\right )&\psi^i_{port}\left ( 0\right )G\left ( \boldsymbol{r},\boldsymbol{r}^i_{port},E_{in} \right ) \\
G\left ( \boldsymbol{r},\boldsymbol{r}^i_{port},E_{in} \right )&= \sum_{n}\frac{\phi_n\left ( \boldsymbol{r}\right )\varphi _n^*\left ( \boldsymbol{r}^i_{port} \right ) }{E_{in} - E_{in}^n}
\end{align}
where $E_{in}^n$ is the eigenvalue of the lattice Hamiltonian $H_{in}$. $\phi^n_{in}\left ( \boldsymbol{r}\right )$ and $\varphi^n _{in}\left ( \boldsymbol{r} \right )$ are the right and left eigenfunction of the lattice Hamiltonian $H_{in}$ defined by
\begin{align}
H_{in} \phi^n_{in}\left ( \boldsymbol{r}\right ) &= E^n_{in}\phi_{in}^n\left ( \boldsymbol{r}\right )  \\
H_{in}^\dagger \varphi ^n_{in}\left ( \boldsymbol{r}\right )&=  \left ( E^n_{in}\right ) ^* \varphi ^n_{in}\left ( \boldsymbol{r}\right )
\end{align}\par
\textit{$\eta $-pseudo-Hermitian condition}---The complete system is $\eta $-pseudo-Hermitian satisfies
\begin{equation}
H_{total}^\dagger = \eta H_{total}\eta^{-1}
\end{equation}
where $\eta$ is a Hermitian invertible linear operator. Using the non-Hermitian indefinite inner-product defined by
\begin{equation}
\left (\varphi _{in}\left ( \boldsymbol{r}\right ),\phi_{in}\left ( \boldsymbol{r}\right )\right )=\int \varphi^* _{in}\left ( \boldsymbol{r}\right ) \eta \phi_{in}\left ( \boldsymbol{r}\right )d\boldsymbol{r}
\end{equation}
and $\eta $-pseudo-Hermitian, one can get
\begin{equation}
\left [ T_{port}^i\left ( w\right )\right ] ^*\psi_{in}\left ( \boldsymbol{r}^i_{port}\right )=\psi^i_{port}\left ( 0\right )
\label{eq:condition}
\end{equation}\par
\textit{$K^T$}---The wavefunction of the i-th port $\psi^i_{port}\left ( \boldsymbol{r}^i_{port}\right )$ far away from perturbed region or in the TEM region can be written as 
\begin{equation}
\psi^i_{port}\left ( z^i_{port}\right )=A^{out}_{port}e^{ikz^i_{port}}+A^{in}_{port}e^{-ikz^i_{port}}
\end{equation}
With the $\eta $-pseudo-Hermitian condition 18 and the Eq.19, the relation 12 becomes
\begin{equation}
A_{in}-iKA_{in}=A_{out}+iKA_{out}
\end{equation}
where 
\begin{equation}
K_{ab}\left ( w\right )=\frac{\left [T^{a}_{port}\left ( w\right ) \right ]^*}{\sqrt{k}}G\left ( \boldsymbol{r}^a_{port},\boldsymbol{r}^b_{port},w\right )\frac{T^{b}_{port}\left ( w\right ) }{\sqrt{k}}
\end{equation}
So, the scatter matrix $S$ can be expressed as
\begin{equation}
S=\frac{A_{port}^{in}}{A_{port}^{out}}=I+\frac{2W^\dagger W}{i\left ( w -H_{in}\right )+WW^\dagger}
\end{equation}
where
\begin{equation}
W_{nm}=\frac{\sqrt{2}T^n_{port}\left ( w\right )\varphi_n ^*\left ( \boldsymbol{r}_{port}^m \right )}{\sqrt{k}}
\end{equation}
Comparing the Eq.\ref{s_parameter} in the main article, we get $K^T_{n,m}= W_{nm}$.

\end{document}